\documentclass[referee]{emulateapj}
\usepackage{epsfig}
\usepackage{amsmath}
\usepackage{natbib}

\begin{document}

\title{Strong Gravitational Lens Inversion: A Bayesian Approach}
\author{Brendon J. Brewer \& Geraint F. Lewis}
\email{brewer@physics.usyd.edu.au}
\email{gfl@physics.usyd.edu.au}
\affiliation{Institute of Astronomy, School of Physics, A28, 
University of Sydney, NSW 2006, Australia}

\begin{abstract}
If an extended source, such as a galaxy, is gravitationally lensed by a massive object in the foreground, the lensing distorts the observed image. It is straightforward to simulate what the observed image would be for a particular lens and source combination. In practice, one observes the lensed image on the sky, but blurred by 
atmospheric and telescopic effects and also contaminated with noise. The question that 
then arises is, given this incomplete data, what combinations of lens mass distribution and source surface brightness profile could plausibly have produced this image? This is a classic example of an inverse problem, and the method for solving it is given by the framework of Bayesian inference. In this paper we demonstrate the application of Bayesian inference to the problem of gravitational lens reconstruction, and illustrate the use of Markov Chain Monte Carlo simulations which can be used when the analytical calculations become too difficult. Previous methods for performing gravitational lens inversion are seen in a new light, as special cases of the general approach presented in this paper. Thus, we are able to answer, at least in principle, lingering questions about the uncertainties in the 
reconstructed source and lens parameters, taking into account all of the data and any prior information we may have.
\end{abstract}


\maketitle

\section{Introduction}
The number of known gravitational lenses has grown steadily over recent decades \citep[e.g.][]{2005A&A...436L..21C}. They hold much promise for being able to constrain theories about dark matter (by probing the mass distribution of the lensing object), and also for being able to magnify distant sources into view, that would have been unobservable otherwise. However, it is slightly disconcerting that as the number of observed gravitational lens systems has increased, so has the number of seemingly distinct methods for analysing them. A few examples are: The Ring Cycle \citep{1989MNRAS.238...43K}, LensCLEAN \citep{2004MNRAS.349....1W}, LensMEM \citep{1996ApJ...465...64W,2005MNRAS.360.1333W}, Semi-linear Inversion \citep{2003ApJ...590..673W}, and Genetic Algorithms \citep{2005PASA...22..128B}. 

These techniques all have their own particular advantages and disadvantages, and it is not clear 
whether or not there is one method that is in some sense more justified than the 
others. In this paper we seek to unify these approaches and show that they are all different methods of doing essentially the same thing. Of course, practical considerations such as computational efficiency must also be taken into account in judging the relative merits of different inversion methods, although this is not such a huge issue due to increasing computer power and the relative rarity of discoveries of these systems. In this paper, we only consider the question of theoretical justification of these methods; discussions of the practical considerations can be found in the respective literature.

\section{Bayesian Inference}
Bayesian Inference is an approach to statistical analysis that is based almost entirely on probability theory, viewed as a generalisation of logic to include uncertainty \citep{1995PhDT.........9L}. This was the original approach to probability theory as developed by Laplace, but was largely rejected in the early 20th century in favour of the ``frequentist'' school of thought, where probabilities can only be interpreted as long-run frequencies in a random experiment. However, Bayesian methods enjoyed a resurgence in the latter half of the 20th century. One reason for this is the proofs of \citet{cox}, who showed that any system which measures plausibility with real numbers, and satisfies several consistency criteria, is equivalent to probability theory. More convincing, however, is the rapidly growing number of practical examples of its use in real applications \citep[e.g.][ and many others]{2002PhRvD..66j3511L, 2004ApJ...610.1213E, 2005PASA...22..153W}.

The product rule of probability is as follows. For two propositions $A$ and $B$, the probability that both $A$ and $B$ are true, given some background information and/or assumptions $I$, is given by

\begin{equation}
P(AB|I) = P(A|I)P(B|AI) = P(B|I)P(A|BI)
\end{equation}

where $P(B|AI)$ is the probability that $B$ is true, {\it given} that A is 
true. Rearranging the above equation gives Bayes' theorem:

\begin{equation}\label{bayes}
P(A|BI) = \frac{P(A|I)}{P(B|I)}P(B|AI)
\end{equation}

By considering propositions of the form ``The quantity X lies in the 
interval $(x,x+dx)$'', it can be shown that a version of Bayes's theorem 
also holds for probability density functions (PDFs) of continuous variables. Thus, whenever you 
want to learn the value of some unknown quantities $\theta$, 
given some other known quantities (the data $D$) that depend on the 
$\theta$ values in a probabilistic (i.e. not entirely predictable) way, then the correct procedure is to calculate the probability distribution of the unknown quantity, given the data:

\begin{equation}\label{bayes2}
p(\theta | DI) \propto p(\theta | I)p(D|\theta I)
\end{equation}

which is the ``posterior'' PDF of the parameters given the data. The $p(D|I)$ term which should be in the denominator has been omitted, because it does not depend on $\theta$ and can be absorbed into the normalisation constant. Thus, Bayes' 
theorem tells us how to modify our state of knowledge of the unknown $\theta$ values by taking the data $D$ into account. However, the posterior PDF depends on $p(\theta | I)$, the 
so called ``prior distribution'' of the unknown quantities. In other 
words, we can't learn from the data unless we specify our prior knowledge 
(or ignorance) about the values of those unknown parameters, before 
observing the data.

The second term in equation~\ref{bayes2}, the probability density of the data if we knew the values of the parameters, is called the sampling distribution if we consider a fixed set of parameters, and imagine calculating the probability of obtaining different data sets $D$. However, for a fixed data set (the one that is observed), and considering its dependence on $\theta$, it is called the likelihood function. It is immediately clear that the ``maximum likelihood estimate'' of statistics is just the mode of the posterior distribution, for a uniform prior distribution, because the posterior PDF is proportional to the likelihood function in this case. As shown by \citet{2002nrc..book.....P}, this leads to a neat justification for the method of least squares, as being appropriate when the sampling distribution is Gaussian.

The presence of the prior term in the Bayesian approach has been the subject of much controversy \citep{1995PhDT.........9L}, and is usually the reason why some people avoid it. The point of the prior distribution is not to describe randomness or variation of the quantity, but our state of knowledge about it. However, the results of a calculation are generally insensitive to the choice of prior, particularly when it is diffuse compared to the likelihood function. In addition, conventional statistics, while appearing more ``objective'' due to the lack of things like prior distributions, has merely swept the matter under the carpet. Use of statistical methods that don't make reference to prior distributions can usually be shown to be equivalent to a specific choice of prior; the maximum likelihood estimate mentioned above is one example. There also exist a limited number (at this time) of general principles for the objective assignment of prior distributions, given particular types of prior information \citep{problogic}. A simple example is the principle of insufficient reason, given by Laplace; if we only know that there are $N$ possibilities that are mutually exclusive and exhaustive, and nothing more, we assign equal prior probabilities of $1/N$ to each.

\section{Gravitational Lensing}
One of the main observable predictions of the theory of general relativity is that light rays can be bent as they pass through the gravitational field of a massive object. Usually, we assume that all of the mass in the lensing object is located in a plane, called the lens plane. If the coordinates on the lens plane are $x$ and $y$, the mass distribution of the lensing object is described by a function $\rho(x,y)$. A light ray passing through the lens plane will be bent by an angle called the deflection angle $\mathbf{\alpha}(x,y)$. For the case of a point mass lens, the deflection angle is directed towards the mass, and has magnitude proportional to $1/r$, where $r$ is the distance from the point mass. For the case of a continuous mass density, the deflection angle at a point $(x,y)$ in the lens plane is obtained by integrating the deflection angles due to all other mass elements. In the usual scaled coordinates, any light that is observed at position $(x,y)$ in the image plane had its origin at the position $(x_s,y_s)$ in the source plane, given by

\begin{equation}\label{scaledlens}
x_s = x - \alpha_x(x,y)
\end{equation} 
\begin{equation}
y_s = y - \alpha_y(x,y)
\end{equation} 

Note that equation~\ref{scaledlens} gives a unique source plane position $(x_s,y_s)$ for a given image plane position $(x,y)$. In general, though, the inverse function does not exist, so any particular position $(x_s,y_s)$ in the source plane can be mapped to multiple positions in the lens plane. This is the mathematical reason why the background source is often multiply imaged in gravitational lens systems. For an extended source, the source may be described by a surface brightness profile $S(x_s,y_s)$. The observed image is usually considered as a function over the lens plane, $I(x,y)$. Since surface brightness (intensity per unit area) is not changed by lensing, we have that the observed image is given by

\begin{equation}
I(x,y) = S(x_s(x,y),y_s(x,y))
 = S(x - \alpha_x(x,y),y - \alpha_y(x,y))
\end{equation}

When the source and image are pixellated, this relation does not hold exactly because the image consists of the integrated surface brightness over the pixels. However, there is a linear relationship between the source and the image in this case, which is given in section~\ref{pixellated}.

\section{Method}
In this section we present the general procedure of the calculations that need to be done, but the equations are left in a generic form, appropriate for whatever parameterisation of lens and source is chosen. Later on, we will specialise to the case of a pixellated source and a parameterised lens model. In the context of lensing, we have an underlying source $S$ and a lens $L$, which we wish to infer from observations, which are the observed image $O$. In other words, we want the joint posterior distribution of the $S$ and $L$ parameters (whatever quantities we are using in our mathematical description of the source and lens), given $O$. Bayes' theorem states

\begin{equation}\label{bayeslens}
p(LS|OI) \propto p(LS|I)p(O|LSI)
\end{equation}

where the factor $p(O|LSI)$ is determined by the model of the noise. If the lens parameters are of interest, but the source is not, then the inference about the lens is given by the marginal probability density of the $L$ parameters:

\begin{equation}
p(L|OI) \propto \int{p(LS|I)p(O|LSI) dS}
\end{equation}

In general, any prior knowledge of the source will not affect our prior knowledge about the lens, and vice versa. Therefore the joint prior PDF for $L$ and $S$ can be factorised into independent priors for $L$ and $S$, giving

\begin{equation}
p(L|OI) \propto p(L|I)\int{p(S|I)p(O|LSI) dS}
\end{equation}

In some special cases, it may be possible to evaluate these integrals analytically. This is not the case in the following discussion, however there are numerical methods which can be used.

\section{Pixellated Sources}

When all is said and done, we would like to describe our knowledge of the source and the lens by a finite number of values, and the uncertainties in those values. In this paper, we are concerned with the case of extended sources, which can have detailed structure in them. Thus, we need to choose a parametrisation that allows for the possibility of arbitrarily complicated structures in the source. Pixels are the obvious choice.

\subsection{What are we doing?}
Suppose we are given an observed image. For any lens model we choose, it is possible to find a source light distribution that reproduces the observed image. In the case that two points with different brightness in the image map to almost the same point in the source plane, we could get around this by having rapidly varying structure in the source, with extremely high positive and negative values of surface brightness. However, nobody would seriously propose such a reconstruction; we already have prior knowledge that rules out this possibility. We expect in advance that the source should be non-negative and smooth on some scale. There are several ways of taking this into account, using pixels is one way, because structure within pixels is impossible. Other possibilities include using non-square basis functions (for example, reconstructing the source as a combination of Gaussian blobs), use of ``fuzzy pixels'', or by choosing a prior distribution that rewards smooth sources.

In using pixels to describe the source light profile, we are restricting our resolution to scales larger than the size of the pixels. It is often clear how to judge in advance the pixel size that is necessary. With gravitational lensing, however, intuition suggests that our image will allow us to reconstruct the source with better resolution in some regions of the source plane than in others. For example, we might expect that we could achieve higher resolution in the reconstructed source for areas in the source plane that are near the caustics - these regions have been imaged more times (ie they affect more pixels of the image) than the others, so our image provides more constraints on them. \citet{2005ApJ...623...31D} have devised a clever method of adaptive pixellation of the source plane for use with their semi-linear inversion method, that puts more pixels on high magnification regions.

In this paper we use a uniform pixel size, and since we calculate the posterior distribution for the brightness of each pixel, will be able to see directly which pixels are strongly constrained by the observations, and which ones aren't. Hence we can check if the high magnification regions of the source plane really are more tightly constrained by the data.

In reality, sources are not made up of pixels, ie square regions of constant surface brightness. We are obliged to use them because we must have some way of describing the source by a finite set of numbers. However, observed images taken with CCDs do actually consist of pixels. The intensity in each pixel corresponds to an estimate of the {\it integral} of the true surface brightness (``intensity density'') of the image, over the area covered by the pixel. Hence, the object we wish to infer (the source) is a surface brightness function $S(x_s,y_s)$, yet our observed data consists only of integrals of that function over certain regions. Thus, our data can only ever constrain those integrals, and prior information is the only way of deciding between different reconstructions that all fit the data.

Bayesian methods give us the posterior PDF of all of the pixel values, given the data. If we reconstruct with a large number of pixels, then any individual pixel is likely to be constrained very weakly by the data, so the marginal posterior PDF of any individual pixel will be very wide. However, a question we would like to answer is ``if the lensing object was not there, and we observed the source, what would we see?''. Since observed images are pixellated, in effect we want to estimate the integral of the surface brightness over some square regions, which may be larger than the pixels we did the reconstruction with. We will see that, while the individual pixels are only weakly constrained by the data, sums of them over particular regions (which approximate the integral of the surface brightness over those regions) can be quite precisely determined.

A strong analogy can be drawn between this reasoning and that used in statistical mechanics. In statistical mechanics, we start off completely ignorant of the microscopic state of the system. Then, measurement of a macroscopic thermodynamic variable narrows down the range of possible ``microstates'' that the system could be in, but there are still a huge number that are compatible with the macroscopic constraints imposed by the data [the ``ensembles'' of statistical mechanics are somewhat like posterior PDFs of the microscopic variables, given the macroscopic data, see \citet{1957PhRv..106..620J}]. When we use this to make predictions about other macroscopic quantities, though, we find that there is enough information to make accurate predictions of these.

\subsection{Underdetermined Problems}
It is well known that, when attempting to infer a large number of parameters from some noisy data, the large number of parameters makes it possible to ``overfit'' the data by fitting the noise. With conventional strategies, the best possible fit within the parameter space is obtained, and is usually accepted or rejected based on the $\chi^2$ criterion \citep{2002nrc..book.....P}. This penalises the use of large numbers of parameters, and behaves like a kind of ``Occam's Razor''. From a Bayesian perspective, however, it is clear that $\chi^2$ is not the final answer to the question of goodness of fit. It is possible that a correct model could give a best fit which is not good enough by the $\chi^2$ criterion, yet the fit is still good over a wide range of parameter space. To reject a model altogether, it is not the value of $\chi^2$ of the best fitting parameters that is important (although it may be a useful indicator), but the ``evidence'' value $P(\mathbf{O}|I)$, which is required in order to test one model against a particular alternative model $I_2$, as can be seen by writing down Bayes' theorem for the posterior probabilities for two competing models $I_1$ and $I_2$.

The answer we get from Bayesian analysis is the joint posterior PDF of all of the parameters, indicating what values are plausible in the light of the data. Even though the {\it best} fitting set of parameters may dramatically overfit the data, the volume of parameter space that is near this peak is exceedingly small, and decreases with the number of parameters. Thus, the overwhelming majority of plausible reconstructions do not ``fit the noise'', even if we use a large number of parameters.

\citet{2003ApJ...590..673W} noted that the use of ``regularisation'' made the $\chi^2$ criterion inapplicable, by effectively reducing the number of free paramaters, by tying some together or limiting their allowed ranges. Since the use of regularisation is equivalent to a particular choice of prior distribution (see section~\ref{relation}), it is possible to compare different reconstructions objectively. The theory of Bayesian model selection \citep{2002MNRAS.335..377H} provides the means for accomplishing this, and automatically displays phenomena such as Occam's Razor, and penalising models that require fine tuning (one model may fit the data well, but only in a small region of its parameter space, while another model which fits the data over a wide range of its parameter space will be preferred). Unfortunately, this usually involves integrals of complicated functions over many dimensions, so is often impractical. With this problem, $\chi^2$ is not strictly applicable, but the alternative is difficult, and is beyond the scope of this paper. It seems as if this problem can become arbitrarily complicated. This is true of every problem in science. Any phenomenon can be modelled either crudely or in great detail, and it is only the limits in the abilities and attention spans of the people studying them that determines when they decide to stop.

\subsection{Gravitational Lensing of Pixellated Sources}\label{pixellated}
When both the image and the source are pixellated, they can be regarded as vectors $\mathbf{I}$ and $\mathbf{s}$ respectively, where each component of the vector is the value of a pixel. In general, these vectors will be of different lengths. Due to conservation of surface brightness, if the source was multiplied by a constant or had a constant added to it, the effect of this on the image, and hence the image pixel values, would be the addition or multiplication of the same constant. In other words, lensing is a linear operator, so for the case of pixellated images and sources, there must be a linear relation between $\mathbf{I}$ and $\mathbf{s}$. ie a matrix $\mathbf{L}$ exists such that

\begin{equation}
\mathbf{I = Ls}
\end{equation}

The form of the matrix $\mathbf{L}$ is completely determined by the lens mass distribution $\rho(x,y)$, and can be calculated by ray tracing  \citep{1994ApJ...426...60W}. The blurring by the point spread function is also a linear operation, so can be represented by another matrix multiplication, alternatively, we regard $\mathbf{L}$ as being the matrix product of a lensing matrix and a blurring matrix. The observed image is vector $\mathbf{O}$ which is related to the true source $\mathbf{s}$ by

\begin{equation}
\mathbf{O = Ls + N}
\end{equation}\label{linear}

where $\mathbf{L}$ is an {\it unknown} lensing/blurring matrix and $\mathbf{N}$ is a vector of ``random noise'', the observational errors in each pixel of the observed image. Thus, this problem is of a standard type \citep{2002nrc..book.....P} as the structure of the problem in equation~\ref{linear} arises in many different contexts, and is not limited to CCD pixel values. Here, the matrix $\mathbf{L}$ is unknown, but we will describe the lens by a parametrised model, so the number of degrees of freedom of the matrix $\mathbf{L}$ is not too large. The point spread function is assumed to be known for our simulations; this is at least approximately true in practice, as the point spread function can be measured or simulated from the instrumental response to a point source. Due to the presence of the blurring, we are effectively also doing Bayesian deconvolution \citep{2004ApJ...610.1213E}.

\subsection{Prior Distributions, Relation to Other Methods}\label{relation}

The choice of prior distributions often gets little attention. However, it is usually the case that neglecting this question by using some other method is equivalent to using a particular prior distribution. For example, it is quite easy to show that the method of \citet{2003ApJ...590..673W}, being a least squares method, is equivalent to finding the peak of the posterior PDF $p(\mathbf{s}|\mathbf{O}LI)$ for a uniform prior distribution $p(S|I) = constant$, and Gaussian noise. This has the obvious drawback that it allows pixels to be negative, and hence ignores the most basic prior information that we know of - surface brightness is a nonnegative quantity. However, this may be a small price to pay for the convenience it delivers. 

A simple way to rectify this is to consider the problem as one of constrained least squares, where the $\chi^2$ deviation between the model and the data is minimised subject to a positivity constraint on all of the source pixels values \citep{2005PASA...22..128B}. Least squares is equivalent to the Bayesian approach with a uniform prior distribution, but with negative values forbidden. There are some problems associated with this method that are discussed in section~\ref{failure}. These problems were not noticed previously because the methods focused on {\it optimisation} of the source (i.e. finding the best fit), rather than {\it exploration}, where the set of all plausible fits is considered.  

When ``regularisation'' is used \citep{2003ApJ...590..673W,2002nrc..book.....P}, whereby an additional term representing ``quality'' is added to the merit function to be optimised, this is equivalent to assuming a particular form of prior distribution, and then seeking the most probable reconstruction (for a given lens model) by maximising $p(\mathbf{s}|\mathbf{O}LI)$. The Maximum Entropy Method \citep{1989MNRAS.238...43K} is similar in nature, and has the added benefit that positivity of the source is guaranteed. These approaches tend to give results that are useful and visually appealing, but some issues arise when these are applied to astronomical images.

Astronomical images are mostly blank. This is a simple and important fact that none of the above priors take into account, but which is important. For example, use of the prior that is constant everywhere where the source pixels are positive, while seeming very conservative about the values of individual pixels, can end up making dogmatic statements about the sum of many pixels. Use of regularisation or the entropic prior of LensMEM actually exacerbate this problem, by making reconstructions which are compressed (have more moderate values and less extreme ones) more likely a priori. In later sections with simulated data, we will use a prior that, while not having as sophisticated a justification as the entropic prior, is closer to what we actually think about astronomical data.

We will make the conventional choice of a Gaussian distribution for the noise, leading to the following sampling distribution/likelihood function $p(\mathbf{O}|\mathbf{s}LI)$,

\begin{equation}\label{gaussian}
p(\mathbf{O}|\mathbf{s}LI) \propto \prod_{i=1}^n[e^{-\frac{1}{2}\left(\frac{O_i - \sum_{j=1}^mL_{ij}s_j}{\sigma_i}\right)^2}] = e^{-\frac{1}{2}\chi^2}
\end{equation}

where the $\sigma_i$ are the ``uncertainties'' of each pixel of the observed image. Before applying any lensing reconstruction procedures to the data, it has usually been subjected to some other procedures, such as foreground galaxy subtraction and also the subtraction of the mean sky brightness. Thus, it is unreasonable to expect to be able to use a Poisson sampling distribution (from photon counting arguments), because after these necessary procedures been performed, the data are no longer integer valued, and can be negative.

As long as the $\sigma_i$ correctly indicate the expected (rms) magnitude of the noise, then the independent Gaussian PDF is the most conservative and general probability assignment we could make. It does not necessarily have to correctly represent the long run frequency distribution of the errors in repeated experiments (if such a thing exists), as discussed by \citet{problogic} and demonstrated empirically by \citet{bretthorst}. As long as the Gaussian model does not assign extremely low probability to the actual errors that are present in the one observed image that we actually have, then no problems will arise from its use. The worst case scenario is that parameter estimates will be slightly more conservative than they would have been if we had additional information about the noise.

\subsection{Prior Information About the Lens}
In order to set up a prior distribution $p(L|I)$ for the parameters of the lens model, we must consider what little information we have about the lens mass distribution, without considering the image. However, we are allowed to consider the image of the foreground galaxy as information which is relevant to the lens model parameters. In the dark matter paradigm, there is typically a poor correlation between the observed light distribution of the foreground lensing galaxy and the way its mass is distributed \citep{2003ApJ...595...29R}. In this case, the prior distributions for the lens parameters would be very diffuse, with the galaxy image only providing information about the general order of magnitude of the lens parameters. 
However, if we assume some particular relativistic theory of MOND [such as \citet{2004PhRvD..70h3509B}] is true, and its implications for lensing are known, then the observed galaxy image will provide quite strong constraints on the parameters of the lensing model. Thus, lensing can provide a test of dark matter vs MOND, because if MOND is used, the brightness profile of the lensing galaxy should all but fix the lens model. In the coming simulations, uniform priors are used for all lens parameters. This has little effect on the conclusions because the data are able to constrain the lens parameters quite strongly.

\section{Markov Chain Monte Carlo}
It is usually straightforward to write down the posterior PDF for all of the unknown parameters (equations~\ref{bayeslens} together with~\ref{gaussian}). However, written in this form it is often not of much use. We are usually interested in some of the parameters, but not all of the others. For example, we might want to know the posterior PDF for the total mass in the lens, so we can quote an estimate and its uncertainty. In principle, the way to deal with this is to obtain the marginal PDF for all of the wanted parameters, by integrating the joint PDF with respect to all of the unwanted parameters. However, in practice this is usually impossible to do analytically.

Markov Chain Monte Carlo (MCMC) methods are a computational tool for solving this problem. They seek to generate a random sample from the posterior PDF [a cosmological example is \citet{2002PhRvD..66j3511L}]. By looking at only one parameter of the models in the sample (for example as a histogram), we can immediately get an idea of the marginal posterior PDF for that parameter. We can also approximate its mean and standard deviation (which we might give as the estimate and the uncertainty) by using the mean and standard deviation of the sample. The Markov Chain aspect comes from the way these methods work. Most work by taking a kind of random walk through the parameter space, with the transition probabilities constructed in such a way that the stationary distribution of the Markov Chain is the posterior distribution that we are interested in. Then, in the long run, the random walk visits different regions of the parameter space in proportion with their posterior probability \citep{mcmc}.

\subsection{The Metropolis Algorithm}
The Metropolis algorithm \citep{metro} is the starting point for many Markov Chain Monte Carlo schemes. Start from a point $\mathbf{x}$ in the parameter space. Denote the target distribution (which we want to sample from, in our case it is the posterior distribution $p(\mathbf{SL}|\mathbf{O}I))$ by $\pi(\mathbf{x})$. A proposal step is made, from the point $\mathbf{x}$ to the point $\mathbf{y}$ according to the {\it proposal distribution} $q(\mathbf{y}|\mathbf{x})$, which is assumed to be symmetric, meaning that $q(\mathbf{y}|\mathbf{x}) = q(\mathbf{x}|\mathbf{y})$. This proposal step is always accepted if $\pi(\mathbf{y}) \ge \pi(\mathbf{x})$, otherwise it is accepted with probability $P = \frac{\pi(\mathbf{y})}{\pi(\mathbf{x})}$. This procedure is repeated many times, resulting in a random walk through the parameter space whose long term frequency distribution is equal to the target distribution. The performance of this algorithm depends on the form of the proposal distribution (the method should always work, but some proposal distributions may explore the parameter space more quickly). Typically, a normal distribution centred at the current value is used, and the variance is adjusted to achieve a moderate acceptance rate of 20-50 per cent.

The success (or otherwise) of Markov Chain Monte Carlo methods is highly dependent on the starting point of the random walk. If the starting point is somewhere of extremely low posterior probability, the chain can wander around and get stuck in local minima, possibly never reaching the correct region. In principle, it eventually will, for instance it might spend $10^9$ years in one local maximum before leaping by chance into the high probability region, where it would spend $10^{20}$ years, say. Unfortunately, there are no general methods for discovering whether this has occurred. MCMC can also struggle to cope with a situation where good fit is achievable within two parts of the parameter space that have very different parameter values. This is a problem shared by virtually all other methods, some more so than others. If multiple maxima are expected to occur then simulated annealing or genetic algorithms may be helpful in finding the maxima. Once the maxima are found using any method, they can be used as a starting point for the MCMC method for evaluating the uncertainties. For the lens modelling, we specified an initial point where the lens parameters were close to the optimal values, and the source was blank.

\section{Application to Simulated Data}
\begin{figure}
\begin{center}
\includegraphics[width = 0.5\textwidth]{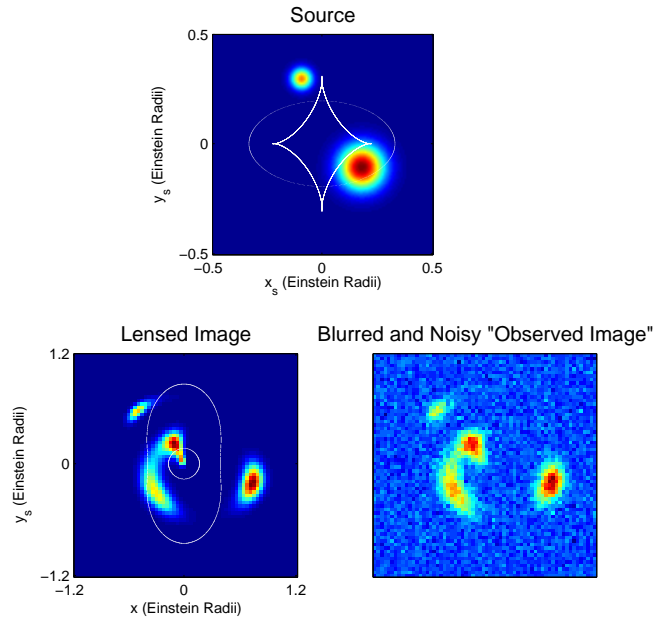}
\end{center}
\caption{Simulated data for testing our method. The source was a double Gaussian, which was lensed via a ray tracing method. The sizes of the regions of the source plane and image plane that are covered by these pictures are the same throughout the paper. The caustics and critical lines are shown on the source plane and the image plane respectively. Throughout the paper, we consider only small changes in the lens parameters, so the caustic and critical line geometries do not change significantly.\label{sim}}
\end{figure}

We tested this algorithm on some simulated data. The source profile was two Gaussians, and these were lensed to create an image on a 64x64 pixel grid. The image was created by ray tracing, with the source evaluated analytically. Hence we can see the effect that assuming pixels has on the final conclusions, when the source isn't really made from pixels. This image was then convolved with a point spread function, and random noise of known standard deviation was also added. These are shown in Figure~\ref{sim}. The units were chosen so that the image covers a region 2.4 Einstein Radii across. The lens model that was used was a Pseudo Isothermal Elliptical Potential (PIEP) lens model \citep{2005PASA...22..128B} with parameters $b = 0.5$, $\epsilon = 0.25$, $r_c = 0.1$, centred at the origin and with the principle axes of the ellipse aligned with the $x$ and $y$ axes. The $b$ parameter describes the strength of the lens, $\epsilon$ is the ellipticity, and $r_c$ is the core radius.

\subsection{The Positive Uniform Prior}\label{failure}
At first, a reconstruction was attempted, using the prior distribution which is a constant as long as all source pixels are positive. The proposal distribution for moving from one step of the chain to the next was done by adding a normally distributed random number to each pixel, and then taking the absolute value. To achieve a moderate acceptance rate in the Metropolis method, the standard deviation of the proposal distribution was set to 5 per cent of the noise level in the image, in the case where all pixels are updated at once. For this run, a 24x24 pixel grid was used, and the lens parameters were held constant at the known true values; this suffices to illustrate the point.

\begin{figure}
\begin{center}
\includegraphics[width = 0.5\textwidth]{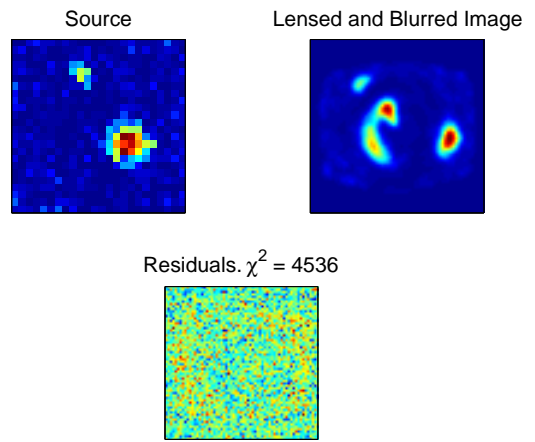}
\end{center}
\caption{A random sample source from the posterior PDF, using the constant prior. Note that the reconstruction is too bright in extended areas, particularly where the source is actually dark, and not multiply imaged (the outer regions).\label{redstuff}}
\end{figure}

A random source sampled from the posterior distribution is shown in Figure~\ref{redstuff}. The corresponding image, and the residuals (difference between the reconstructed and observed image) are also displayed. The $\chi^2$ deviation between the model and the observed image is 4536, which is much higher than the expected value of $64^2 = 4096$ if this model was correct. This is because the reconstructed source is much too bright in just those areas where it should be dark.

This is a fatal flaw of this prior distribution - when it is used, the majority of ``plausible'' reconstructions don't actually match the data! There are several reasons why this effect was not noticed in previous studies. Firstly, they focused on optimisation rather than exploration. The source that maximises the posterior PDF would surely fit the data, but the volume of parameter space around this source is very small due to the positivity cutoffs. Also, some authors have chosen to put a ``mask'' around the image, so that the blank regions are deemed irrelevant and not included in the calculations \citep{2005MNRAS.360.1333W}. Thus, they only looked at the residuals in the bright regions of the image, which is the region that isn't affected by this problem. By masking out blank parts of the image, possibly important information is also being wasted. For example, suppose a model was found that reproduces the bright parts of the image faithfully, but also predicts that there should be bright regions where in fact darkness has been observed. This model would be acceptable if the mask was used, but if we always use the whole image, this situation can never arise.

When the lens parameters were also allowed to vary, the estimates obtained were also slightly incorrect, with the true values lying several standard deviations away from the estimates. Thus, using a positive constant prior (or something that turns out to be equivalent to it, such as constrained least squares) in this situation can bias the lens model results slightly. The biasing effect of the uniform prior increases with the number of pixels that are used, since for more pixels, the implied prior estimate on the integrated flux over any region becomes higher.

\subsection{Two Reconstructions}
The idea behind the prior distributions that were used in the remainder of the runs was as follows. Each pixel has an independent distribution, with some positive probability of being very close to zero. This accounts for the ``astronomical images have a lot of blank regions'' observation noted earlier. If a pixel is not blank, then we at least know the order of magnitude of the surface brightness distribution, so the positive part of the prior distribution was taken as an exponential with a mean value of 100 (the typical brightness scale of the bright parts of the image). The prior distributions will be written down explicitly in later sections. A more sophisticated analysis leads to a slightly different prior which has been called ``Massive Inference'' \citep{1998mebm.conf....1S}, which is a generalisation of maximum entropy, that has similar features to our priors. In the following sections we show how priors based on these seemingly trivial observations lead to more sensible reconstructions.

In all of the following simulations, the lens parameters were also varied. Since this is a more computationally intensive task, involving ray tracing, it was only done once every 10th step. Unfortunately, to achieve a moderate acceptance rate, the proposal distribution for the lens model had to be quite narrow. We used a normal distribution centred at the current lens parameters, with a standard deviation of 0.001. For the first 50,000 steps of the chain, only the source was varied, so that the reconstruction is reasonable before the lens parameters start changing.

\subsubsection{Not Many Pixels}
In this section we present two different reconstructions of the same simulated data, which both lead to sensible results. At first, we reconstructed with a 16x16 pixel model for the source, covering a square region from -0.5 to 0.5 Einstein radii in the source plane. The prior probability density on each pixel was taken as a mixture of two exponentials, with 50 per cent weighting given to each:

\begin{equation}
p(s_i|I_1) = \frac{1}{2}exp(-s_i/2)/2 + \frac{1}{2}exp(-s_i/100)/100
\end{equation}

\begin{figure}
\begin{center}
\includegraphics[width = 0.5\textwidth]{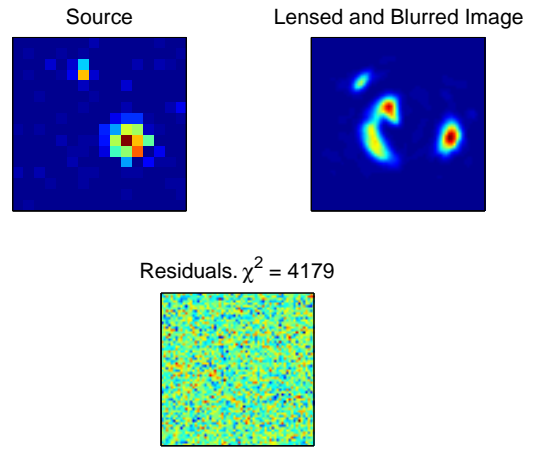}
\end{center}
\caption{A random sample source from the posterior PDF, using the mixture of exponentials prior. The residuals appear as a noise map, without the overly bright regions that occurred with the uniform prior.\label{sample}}
\end{figure}

The first exponential gives a 50 per cent prior probability of the pixel being quite dark, while the other exponential allows for the possibility that the pixel is quite bright. The MCMC approach that was used for this reconstruction was a slightly modified version of the Metropolis algorithm. The prior distribution was actually incorporated within the proposal step, rather than in calculating the ratio of the posterior probabilities. In other words, the proposal step was created such that it would have sampled from the prior distribution, in the absence of any data. Then the acceptance/rejection decision was based on just the ratio of the likelihoods, as the prior had already been taken into account. Figure~\ref{sample} shows a sample reconstruction from the posterior distribution, both the source and the resulting blurred image. The problem of overly bright areas in the reconstruction has disappeared, and the residuals look just like a uniform noise map, as they should.

\begin{figure}
\begin{center}
\includegraphics[width = 0.5\textwidth]{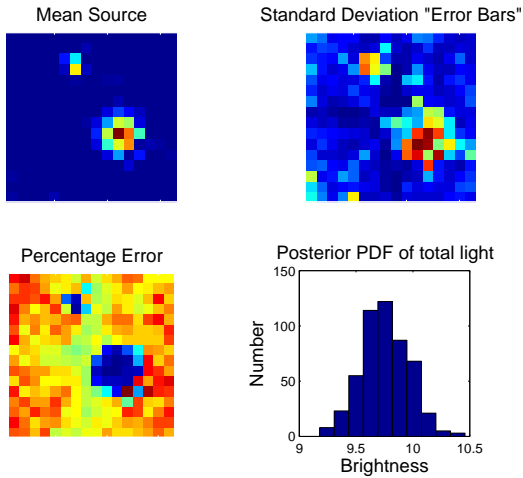}
\end{center}
\caption{Reconstructions with the 16x16 pixel grid, and the prior distribution which allowed a significant chance for each pixel to be dark. The uncertainty of each pixel (taken as the standard deviation of the marginal posterior distribution) is largest for those pixels that are bright, and is smallest for the pixels that are highly magnified. This can be seen from the presence of the diamond shaped dark region in the uncertainty map - this is precisely the high magnification region of the favoured lens model.\label{rec1}}
\end{figure}

The mean and standard deviation of the entire sample are also shown (Figure~\ref{rec1}), as is the posterior PDF of the total amount of light in the source. As described in the introduction, the marginal posterior PDF of each pixel is quite wide, but the inference about a ``macro'' quantity (the total integrated intensity) is fairly strongly constrained at $9.75 \pm 0.21$ (posterior mean $\pm$ standard deviation, so a ``1-sigma'' uncertainty). The reason for displaying the mean of all of the reconstructions is because expectation values satisfy a linear property, so if we want to predict any integral property of the source (for instance, the sum of all pixels within a certain region), the mean of the sum is equal to the sum of the means.

\begin{figure}
\begin{center}
\includegraphics[width = 0.5\textwidth]{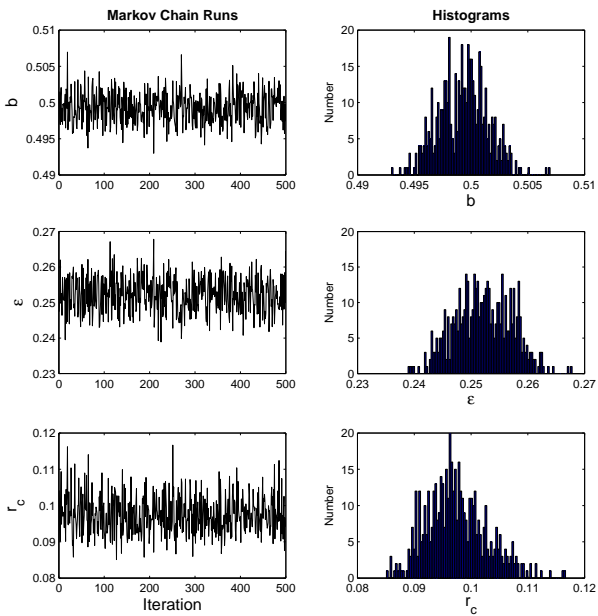}

\end{center}
\caption{Lens model parameters with the 16x16 pixel grid. With this particular pixellation and prior distribution for the source, the lens parameters are quite accurately determined (the true values are 0.5, 0.25 and 0.1 respectively).\label{lens1}}
\end{figure}

The trajectory of the lens model parameters, and the corresponding histograms, are shown in Figure~\ref{lens1}. Only every 100,000th step of the chain was saved, so that there are no strong correlations between one point and the next. We can see from the histograms what the inference would be about the three ``interesting'' lens parameters, $b$, $\epsilon$ and $r_c$. The position of the centre of the lens, and the angle of its orientation, were also varied and the marginal PDFs of those parameters can also be viewed, but since these are uninteresting nuisance parameters, we have not displayed them here.

With a prior distribution on each pixel that allows a significant probability of the pixel being very dark, the inference about the lens parameters is more accurate than with the uniform positive prior. The measured values of $b$, $\epsilon$ and $r_c$ would be $0.4993 \pm 0.0022$, $0.2521 \pm 0.005$ and $0.0975 \pm 0.0053$, respectively, which are consistent with the known true values of 0.5, 0.25 and 0.1. A similar consistency is also seen with the position of the centre of the lens, and its orientation angle, for which the true values were zero.

\subsubsection{Many Pixels}

An alternative pixellated model was also attempted, with 64x64 pixels. With this reconstruction, we used a more informative prior, this time with a prior probability of 90 per cent that the pixel is exactly zero:

\begin{equation}
p(s_i|I_2) = \frac{9}{10}\delta(s_i) + \frac{1}{10}exp(-s_i/100)/100
\end{equation}

To incorporate this in the Metropolis method, we proceeded as before by defining the proposal step so that it would have sampled from the prior distribution if there wasn't any data. In each proposal step, one pixel was selected at random, and switched off with probability 0.9, or on with probablility 0.1, in which case its surface brightness was taken from an exponential distribution with a mean of 100. The accept/reject decision was then based on the likelihood ratio of the proposed step to the current one.

\begin{figure}
\begin{center}
\includegraphics[width = 0.5\textwidth]{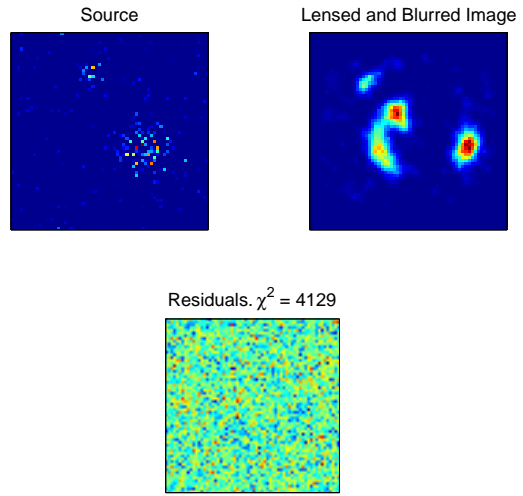}
\end{center}
\caption{A random sample source from the posterior PDF, using the 64x64 pixel grid for the source, and the 90 per cent dark prior. The residuals appear like a noise map, as expected.\label{sample2}}
\end{figure}

\begin{figure}
\begin{center}
\includegraphics[width = 0.5\textwidth]{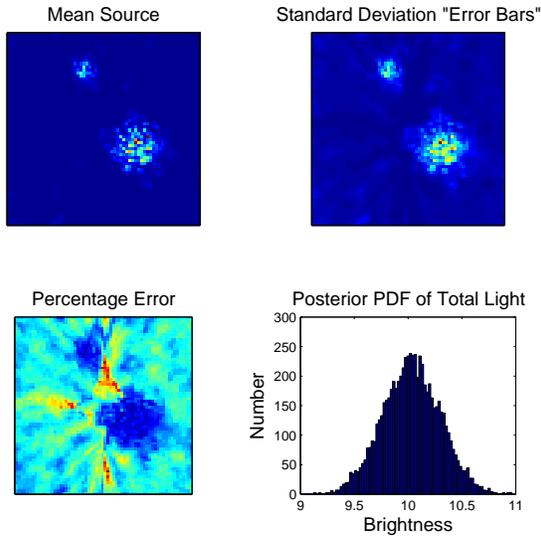}
\end{center}
\caption{Reconstructions with the 64x64 pixel grid, and the prior distribution which allowed a 90 per cent prior probability of a zero in any pixel. The magnification pattern is again present in the uncertainty maps - high magnification regions have low uncertainties, as expected.\label{rec2}}
\end{figure}

\begin{figure}
\begin{center}
\includegraphics[width = 0.5\textwidth]{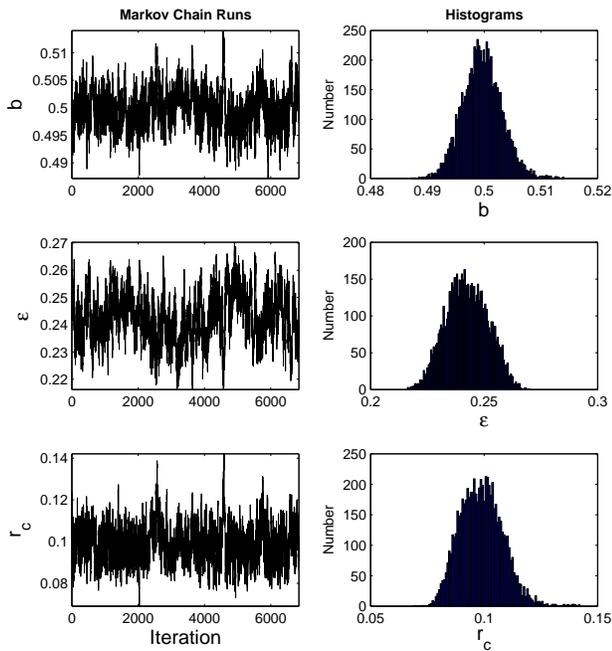}
\end{center}
\caption{Lens model parameters with the 64x64 pixel grid. Due to the increased flexibility that the source is allowed to have with this parameterisation, the lens model parameters are less strongly constrained, with the uncertainties increased by a factor of $\sim$ 3. However, the measurements are still accurate enough for most purposes.\label{lens2}}
\end{figure}

Usually, if a proper model selection analysis is done (by calculating $P(I_2|\mathbf{O})/P(I_1|\mathbf{O})$), the 16x16 reconstruction would be favoured because it has managed to explain the data adequately with fewer parameters. Within the Bayesian framework, however, the more informative prior distribution in the 64x64 case counteracts this tendency to some extent (in principle this can be calculated, but it is beyond the scope of this paper). This is because the posterior probability of a model depends not only on the number of parameters, but also on the amount of prior information that there is about the values of the parameters. A model is favoured if it predicts the observed data (by assigning high probability to it) - but if we do not know much about the values of parameters, then we know less about what the model actually predicts. Hence it must spread its share of probability over a larger range of possible data sets, and therefore assigns lower probability to the data set that was actually observed.

From the error maps in Figure~\ref{rec2}, it is clear that the uncertainty as to the value of a pixel is far from uniform over the source plane. The standard deviation image shows that the uncertainty is greatest in the bright regions. There is also a slight effect due to the magnification pattern, as the diamond-shaped caustic shows up weakly as a region of lower uncertainty. However, the fractional uncertainty (standard deviation divided by the mean) actually follows the opposite trend. The bright values have low percentage error, and within the high magnification regions, the percentage error is highest. This suggests that the answer as to which regions of the source plane are more constrained is not so clear cut.

\section{Conclusions}
Various methods have been proposed for analysing extended gravitational lens systems, in order to extract as much information as possible about the lens and the source. Most have been found to be useful, but all have limitations. One feature common to all previous methods is the lack of a simple way to evaluate the uncertainties in all of the inferred  quantities. By using Bayesian inference, it is possible to reinterpret these methods, showing that they are all equivalent to Bayesian inference with different priors. This suggests how methods may be improved by using more informative priors, and also how uncertainties should be quantified by exploring, rather than maximising, the resulting probability distribution. 
In this paper, we used a simple Markov Chain Monte Carlo method to explore the parameter space and sample from the posterior distribution, leading to quantitative uncertainties on all of the lens model parameters, and also the source pixel brightnesses. Simulated data was used, but we hope to apply this method to actual data in the near future.

When modelling the Einstein ring ER 0047-2808 using the Maximum Entropy Method, \citet{2005MNRAS.360.1333W} showed that is usually possible to find a plausible reconstruction of an image with several different lens models, when only the best fit parameters are considered. Thus, the lens models remain on an even footing and we cannot distinguish which one is more realistic. Luckily, many conclusions about the source are unchanged by the use of different lens models. However, it should be possible to do Bayesian model selection analysis which may show that one of the models is preferred because it fits the data over a larger volume of its parameter space. This possibility will be explored in future contributions, along with an application of these ideas to ER 0047-2808 and other extended gravitational lens systems.

\section*{Acknowledgments}
BJB thanks Tom Loredo and Sarah Bridle for some tips regarding MCMC, and also the referee for helpful comments. GFL acknowledges support through the grant ARC DP 0452234.

\label{lastpage}

\end{document}